\documentclass[reprint,onecolumn,amsmath, amssymb]{revtex4-2}   
\usepackage[utf8]{inputenc}
\usepackage{mwe,float}
\usepackage{caption,subcaption, url}
\usepackage{natbib}
\usepackage[nointegrals]{wasysym}
\usepackage{rotating}
\usepackage[pdftex]{color}
\usepackage{array,tabu,multirow}
\usepackage{graphicx,bm}
\usepackage{dcolumn}
\usepackage{floatpag}  
\rotfloatpagestyle{empty}
\usepackage{amsmath}
\usepackage{amsthm}
\usepackage{amssymb}
\usepackage{latexsym,upgreek}
\usepackage{hyperref,rotating}
\usepackage{xpatch,bigints}
\usepackage{booktabs,subcaption}
\usepackage[dvipsnames]{xcolor}
\usepackage{ulem}
\usepackage{textgreek}
\usepackage{mathrsfs}
\makeatletter
\AtBeginDocument{\xpatchcmd{\@thm}{\thm@headpunct{.}}{\thm@headpunct{\enskip}}{}{}}
\makeatother 
\usepackage{gensymb}
\usepackage{mathdots}

\newcommand{\qand}{\quad \mbox{and} \quad}

\newcommand{\qor}{\quad \mbox{or} \quad}

\usepackage[version=4]{mhchem}

\usepackage{graphicx,kecpm,slashed}
\usepackage{makeidx}
\makeindex

\newcommand{\grad}{\nabla}

\newcommand{\bex}{\begin{example}}
\newcommand{\eex}{\end{example}}
\newcommand{\besp}{\begin{split}}
\newcommand{\ensp}{\end{split}}
\newcommand{\ph}{\phantom}

\newcommand{\La}{\Lambda}

\newcommand{\thet}{\theta}

\newcommand{\om}{\omega}

\newcommand{\by}{\times}

\newcommand{\ovl}{\overline}

\newcommand{\bos}{\boldsymbol}

\newcommand{\btab}{\begin{tabular}}
\newcommand{\etab}{\end{tabular}}
\newcommand{\barr}{\begin{array}}
\newcommand{\earr}{\end{array}}
\newcommand{\bpm}{\begin{pmatrix}}
\newcommand{\epm}{\end{pmatrix}}
\newcommand{\bit}{\begin{itemize}}
\newcommand{\eit}{\end{itemize}}
\newcommand{\ben}{\begin{enumerate}}
\newcommand{\een}{\end{enumerate}}
\newcommand{\bct}{\begin{center}}
\newcommand{\ect}{\end{center}}

\newcommand{\ra}{\rangle}
\newcommand{\la}{\langle}
\newcommand{\bes}{\begin{split}}
\newcommand{\ens}{\end{split}}

\newcommand{\lt}{\left}
\newcommand{\rt}{\right}

\begin{document}
\title{Tensor gauge fields 
and dark matter in 
general relativity with fermions}

\author{Kevin Cahill}
\email{cahill@unm.edu}

\affiliation{Department of Physics and Astronomy\\
University of New Mexico\\
Albuquerque, New Mexico 87106}
\date{\today}

\nopagebreak

\begin{abstract}
The action of
general relativity with fermions 
has two independent symmetries: 
general coordinate invariance and 
local Lorentz invariance.
General coordinate 
transformations 
act on coordinates and tensor indices,
while local Lorentz 
transformations
act on Dirac and
Lorentz indices, 
much like a noncompact 
internal symmetry.
The internal-symmetry 
character of 
local Lorentz invariance 
suggests that it might be
implemented by 
tensor gauge fields 
with their own 
Yang-Mills action 
rather than by 
the spin connection
as in standard formulations.
But because the Lorentz group
is noncompact, 
their Yang-Mills action
must be modified by 
a neutral vector field whose
average value 
at low temperatures is timelike.
This vector field and
the tensor gauge fields 
are neutral and interact 
gravitationally, so they 
contribute to 
hot and cold dark matter.
The two independent symmetries
of the action are reduced 
to a single symmetry 
of the vacuum, local 
Lorentz invariance,  
by the nonzero
average values
of the tetrads 
$c^a_{\phantom{a} k}$.
The local Lorentz 
invariance of
general relativity with fermions 
can be extended to 
local U(2,2) invariance.
If the contracted squares of 
the covariant derivatives  
of the tetrads  
multiplied by the square 
of a mass $M$
are added to the action,
then in the limit 
$M^2 \to \infty$,
the equation of motion 
of the tensor gauge fields
is the vanishing of the 
covariant derivatives 
of the tetrads, which is
Cartan's first equation of structure.
In the same limit,
the tensor gauge fields 
approach
the spin connection.
\end{abstract}

\maketitle

\section{Introduction
\label{Introduction sec}}

The action of 
general relativity with fermions has two 
independent symmetries:
general coordinate invariance
and local Lorentz invariance.
These symmetries are  traditionally implemented
by Cartan's tetrads 
$c^a_{\ph{a} i}$
and by 
the spin connection $\om_i$
which is a quartic polynomial
in the tetrads and their 
first derivatives. 
In this paper,
the spin connection
is replaced by 
tensor gauge fields
with their own 
Yang-Mills action.
\par
General coordinate invariance 
is the defining symmetry 
of Einstein's 
general relativity.
A general coordinate transformation
$x \to x'$
acts on coordinates $x^i$
and on tensor indices $i, k$
but leaves Dirac indices $\a, \b$
and Lorentz indices $ a, b$
unchanged
\begin{align}
\psi'_\a(x') ={} 
\psi_\a(x),
\label {General coordinate invariance}
\quad c_a^{\prime \ph{a} i}(x')  ={}  
\frac{\p x'^i}{\p x^k} \,
c_a^{\ph{a} k}(x) . 
\end{align} 
General coordinate invariance 
is implemented by 
Cartan's tetrads
$c^a_{\phantom{a} k}$ 
and their derivatives.
\par
Local Lorentz transformations
act on Dirac and
Lorentz indices
but leave
coordinates and tensors
unchanged
\begin{equation}
\begin{split}
\psi'_\a(x) ={} &
D_{\a \b}(\La(x)) \, \psi_\b(x)
\\
(D_i\psi)'_\a ={}&
(\p_i + \om'_i )^{\a\b} \psi'_\b
= D_{\a \b}(\La) \, 
(D_i \psi)_\b
\\
c'^a_{\ph{' a} i}(x)  ={}&  
\La^{a}_{\ph{a} b}(x) \,
c^b_{\phantom{b} i}(x)  . 
\label {Local Lorentz invariance}
\end{split} 
\end{equation} 
Local Lorentz invariance
is implemented by the
spin connection $\om_i$
in standard 
formulations~\citep{Utiyama:1956sy,
Kibble:1961ba,
doi:10.1063/1.1703923,
Weinberg:1972tetrads,
Deser:1976ay, Cahill525, GSW274v2}.
\par 
Invariance under 
general coordinate 
transformations 
and invariance under
local Lorentz transformations 
are both 
exact and independent 
symmetries of the action
of general relativity with
fermions.
But while general coordinate 
transformations 
(\ref{General coordinate invariance}) act on 
coordinates and tensor 
indices, local Lorentz transformations
(\ref{Local Lorentz invariance})
act on Lorentz and Dirac
indices leaving coordinates
unchanged.
In this respect,
local Lorentz invariance 
is like a noncompact internal 
symmetry~\citep{Cahill:1978ps,*Cahill:1979qt,*Cahill:1980,*Cahill:1981rq}.
\par
These observations motivated
an attempt~\citep{Cahill:2020lry} 
to implement 
local Lorentz invariance 
by means of 
tensor gauge fields
$L^{ a b }_{\ph{a b} i}$, 
but the focus was 
so exclusively upon fermions
and the spin connection,
that the invariance
of the action of the
tensor gauge fields
was neglected
and repaired belatedly in
an erratum~\citep{Cahill:2022err}.
In that erratum,
a hermitian matrix $h(x)$
was introduced that under a local
Lorentz transformation $\La(x)$
transforms as 
$h'(x) = D^\dag(\La(x)) h(x) D(\La(x))$
in which $D(\La)$ is Dirac's
$D^{(1/2,0)} \oplus D^{(0, 1/2)}$
representation of $SO(3,1)$.
\par
In the present paper,
local Lorentz invariance 
is implemented
by means of 
tensor gauge fields
$L^{ a b }_{\ph{a b} i}$ 
and a real vector field $K_i$
in terms of which the matrix $h$
is realized as 
$h ={} i \b \c^a c_a^{\ph{a} i} K_i
= i \b \c^a  K_a $.
The spin connection
$\om_i ={} 
\teighth
\om^{ a b }_{\ph{a b} i} 
\bos [ \c_a, \c_b \bos ]$
is replaced by a
``Lorentz connection''
$
L_i ={}
\teighth
L^{ a b }_{\ph{a b} i} 
\bos [ \c_a, \c_b \bos ]
$
with field-strength 
$F_{i k} ={} 
\bos [ \p_i +   L_i, \p_k 
+   L_k \bos ] $
and Yang-Mills-like 
action
\begin{equation}
S_L = {} - \frac{1}{4 m^2 \l^2}\int
\tr \lt(F^\dag_{ik} \, h
\, F^{ik} \, \b h \b \rt) 
\sqrt{g} \, d^4x 
\label {S_L}
\end{equation}
in which 
$\b = i \c^0$,
$g = |\det (g_{ik})|$, and
$\l$ is a coupling constant.
The vector field $K_i(x)$ 
makes the trace in the
action $S_L$ invariant
under noncompact Lorentz
transformations, but
the squares of the time
derivatives 
$(\dot L^{ a b }_{\ph{a b} i} )^2$
appear in 
$S_L$ with positive signs
only if the average value
of $K_i$ is timelike.

\par
The action of 
the vector boson $K_i$ is
\begin{equation}
\begin{split}
S_K ={}& 
\int \Big[
- \tfourth 
(D_i K_k - D_k K_i)
(D^i K^k - D^k K^i)
- \tfourth \xi^2 (K_iK^i + m^2)^2 
\Big] \sqrt{g} \, d^4x  
\label {S_K}
\end{split}
\end{equation}
in which $\xi >0$ is a positive
coupling constant.
The potential energy density
$\tfourth \xi^2 (K_iK^i + m^2)^2 $
makes the average value
$ \la 0 | K_i(x) | 0 \ra $
of $K_i(x)$ timelike at 
low temperatures.
\par
At low temperatures,
$ T \ll \xi m $, the
vibrations about
$ \la 0 | K_i(x) | 0 \ra $ 
are massless and neutral, and 
so would 
contribute to hot dark matter.
But at high temperatures,
$ T \gg \xi m $, the 
the field $K_i$ radiates
particles of mass
$m_K = \xi m$
which would contribute to
cold dark matter
low temperatures.
\par
The tensor gauge fields
$L^{ a b }_{\ph{a b} i}$ 
are neutral and massless
and so like the vector field $K_i$ 
would contribute
to hot dark matter.
\par
Since the tensor gauge fields
$L^{ a b }_{\ph{a b} i}$ 
and the vector boson $K_i$
interact with gravitational 
strength, they would have decoupled
much earlier than the photons,
and so would not have been heated
by the annihilations of quarks
and leptons.  They would 
have a present temperature
much colder than
the 2.7 K of the CMB, 
which decoupled
when the present universe
was 380,000 years old and
had a temperature of about 
0.26 eV.
\par
In terms of the connection
$L_i$ and the gauge fields
of the standard model 
$A_i ={} 
i A^\a_{is} \, t^{\a s}$,
the covariant derivative 
of a Dirac field $\psi$ is
defined in this paper as
\begin{equation}
D_i \psi ={}
\big(\p_i +   L_i + A_i
\big) \psi 
\label {Dirac covariant derivative intro}
\end{equation}
rather than in terms 
of the spin connection
as
\begin{equation}
D_i \psi ={}
\big(\p_i + \om_i + A_i
\big) \psi  
\end{equation}
in which
$
\om_i ={} 
\teighth
\om^{ a b }_{\ph{a b} i} 
\bos [ \c_a, \c_b \bos ]
$
and~\citep{Utiyama:1956sy,
Kibble:1961ba,
doi:10.1063/1.1703923,
Weinberg:1972tetrads,
Deser:1976ay, Cahill525, GSW274v2}
\begin{align}
\om^{ab}_{\ph{ab} i} ={}&
\thalf
c^{aj} \lt( \p_i c^b_{\ph{b} j}
- \p_j c^b_{\ph{b} i} \rt)
-
\thalf  c^{bj}
\lt(\p_i c^a_{\ph{a} j} - \p_j c^a_{\ph{a} i} \rt)
 -
\thalf  c^{ak} c^{b \ell}
c^c_{\ph{c} i} 
\lt( \p_k c_{c \ell} -
\p_\ell c_{ck} \rt) .
\label{spin connection}
\end{align}

\par
The present formalism 
has a serious
disadvantage:
it introduces 
one vector boson $K_i$
and
six tensor gauge fields 
$L^{ a b }_{\ph{a b} i}$.
On the other hand, it
has three advantages:
\begin{enumerate}
\item It treats 
local Lorentz invariance as
an internal symmetry
and gives it a Yang-Mills action. 
\item 
It adds to the usual
internal-symmetry
gauge fields
$A_i ={} i A^\a_{is} \, t^{\a s}$ 
in the Dirac covariant derivative
$ D_i \psi ={}
\big(\p_i +   L_i + A_i
\big) \psi $
a linear combination of gauge fields 
$
L_i ={}
\teighth
L^{ a b }_{\ph{a b} i} 
\bos [ \c_a, \c_b \bos ]
$ 
and not a quartic polynomial 
$ \om_i$ in the tetrads
and their derivatives.  
\item At high temperatures,
the neutral vector field $K_i$
radiates particles of mass $\xi m$ 
which at low temperatures 
would contribute to 
cold dark matter. 
\end{enumerate}

\par

The action $S_L$ 
is discussed in 
Sec.~\ref{Action of the Gauge Bosons sec}.
The matrix $h$,
the vector $K_i$, and
the positive signs of 
squares of the time
derivatives 
$(\dot L^{ a b }_{\ph{a b} i} )^2$
are discussed in 
Sec.~\ref{The Matrix h and the Vector K}.
The Dirac action 
\begin{equation}
S_D 
={}
- 
\int  \bar \psi \, \c^a
c_a^{\ph{a} i} D_i \psi 
\,  \sqrt{g} \, d^4x 
\label {Dirac action}
\end{equation} 
in which $D_i$ 
is the covariant
derivative 
(\ref{Dirac covariant derivative intro})
is discussed 
in Sec.~\ref{Dirac Action}.
The actions $S_L$, $S_K$
and $S_D$ are invariant under 
local Lorentz transformations
and under
independent general
coordinate transformations.
 
\par
Although 
general coordinate invariance
and
local Lorentz invariance
are independent symmetries 
of the action,
they are not 
independent symmetries 
of the ground state
of the universe
because they
do not leave invariant
the nonzero average values
of Cartan's tetrads
$  c^a_{\ph{a} k}  $.
Their average values
$ \la 0 | c^a_{\ph{a} k} | 0 \ra$
or
$\tr ( \rho \, c_{\ph{a} i}^{a}(x) )$
reduce the symmetries
of the action ---
general coordinate invariance
and local Lorentz invariance
--- to a single symmetry 
of the ground state:
local Lorentz invariance.
Since nonzero average 
tetrad values
are intrinsic to the theory,  
this reduction of symmetry 
is intrinsic rather than spontaneous.
It is discussed in
Sec.~\ref{Intrinsic Reduction of Symmetry}.

\par
Local invariance under the Lorentz
group SO(3,1) is
extended to U(2,2) 
in Sec.\ref{Is the gauge group U(2,2)? sec}.

\par
If tensor gauge fields do gauge
SO(3,1) 
and if all invariant terms 
of dimension 
$(\text{mass})^4$ 
or less
occur in the action
(which is not obvious in a 
theory of gravity), 
then the total action
would include the contracted squares 
\begin{equation}
S_C 
={} -
M^2 \int D_i c^a_{\ph{a} k} 
\, D^i c_a^{\ph{a} k} 
\, \sqrt{g} \, d^4x  
\label{contracted square of covariant derivative}
\end{equation} 
of the covariant derivatives
of the tetrads
\begin{equation}
D_i c^a_{\ph{a} k} ={}
\p_i c^a_{\ph{a} k} +
  L_{i \ph{a} b}^{\ph{i} a}
\, c^b_{\ph{a} k} - 
\Gamma^\ell_{\ph{\ell} k i} \, 
c^a_{\ph{a} \ell} .
\label{covariant derivative of tetrad}
\end{equation} 
The factor $M^2$ is required
because tetrads are dimensionless.
The linear combination
$L_{i \ph{a} b}^{\ph{i} a}
\, c^b_{\ph{a} k} - 
\Gamma^\ell_{\ph{\ell} k i} \, 
c^a_{\ph{a} \ell}$
of the
tensor gauge field
and the Levi-Civita connection
acquires a mass of order $M$.
In the limit $M \to \infty$
the equation of motion of 
the tensor gauge fields 
$L^{ a b }_{\ph{a b} i}$ arises
from the term $S_C$ and is
$
D_i c^a_{\ph{a} k} 
={} 0 $
which is 
Cartan's first equation of structure.
In the same limit,
the tensor gauge fields approach
the spin connection
as described in 
Sec.~\ref{Cartan's first equation of structure}.

\section{Action of 
Tensor Gauge Bosons}
\label {Action of the Gauge Bosons sec}

The action  
(\ref{S_L}) 
proposed for the 
tensor gauge fields is
\begin{equation}
S_L = {} - \frac{1}{4 m^2 \l^2}\int
\tr  \lt(  F^\dag_{ik}  \, h \, 
\, F^{ik} \, \b h \b \rt) 
\sqrt{g} \, d^4x 
\label 
{action of the Lorentz connection}
\end{equation}
in which
the field strength $F_{ik}$ is \begin{equation}
F_{i k} ={}  
\bos [ \p_i +   L_i, 
\p_k +   L_k \bos ] \, ,
\label{field strength of spin connection}
\end{equation}
the matrix of 
tensor gauge fields 
$L_i$ is
\begin{equation}
\begin{split}
L_i ={}& 
 \teighth \, 
L^{ a b }_{\ph{a b} i}
\, \bos [ 
\gamma_a, \gamma_b \bos ] \, ,
\end{split}
\end{equation}
$h = {} i \b \c^a c_a^{\ph{a} i} K_i$ 
is a $4\by4$ 
hermitian matrix, 
 $K_i$ is a real vector field,
and
$\l$ is a coupling constant.
 \par
 The action 
 (\ref{action of the Lorentz connection})
 is real because
 \begin{equation}
 \begin{split}
\tr  \big(  F^\dag_{ik} \,  h 
\, F^{ik} \, \b h \b \big)^*
={}&
\tr  \big( \b h \b \, F^{\dag ik} \, h 
\, F_{ik} \big)
={}
\tr  \big( F^\dag_{ik} \,  h 
\, F^{ik} \, \b h \b \big). 
 \end{split} 
 \end{equation} 
 \par
The gamma matrices
\begin{equation}
\begin{split}
\c^0 = {}& 
-i \, \bpm
0 & 1 \\
1 & 0 \\
\epm
,
\qquad\qquad
\b ={}
i \c^0
=
\begin{pmatrix}
0 & 1 \\
1 & 0 \\
\end{pmatrix} 
, \qquad
i \b \c^0 ={}
\begin{pmatrix}
1 & 0 \\
0 & 1 
\end{pmatrix} 
\\
{}
\c^i ={}& - i \, 
\bpm
0 & \s^i \\
{} - \s^i & 0 \\
\epm 
, \qquad 
i \b \bos \c =
\begin{pmatrix}
- \bos \s & 0 \\
0 & \bos \s 
\end{pmatrix} 
, \qquad \,\,\,
\c^5 =
\begin{pmatrix}
1 & 0 \\
0 & -1 \\
\end{pmatrix} 
\label {Weinberg's Dirac matrices} 
\end{split} 
\end{equation} 
satisfy $\{ \c^a, \c^b \} 
= 2 \eta^{ab} I$.
The commutators 
$\bos{[} 
\gamma_a, \gamma_b \bos ]$ 
in 
$L_i ={} - \teighth 
L^{ a b }_{\ph{a b} i}
\, \bos{[} 
\gamma_a, \gamma_b \bos ] $
are for spatial $a,b,c=1,2,3$
\begin{equation}
\boldsymbol{[} 
\gamma_a, \gamma_b \bos ]
= 2i \ep_{abc} \s^c I
\qand
\bos[ \c_0, \c_a \bos]
= -2 \s^a \c^5.
\end{equation}
\par
The gauge fields
associated with
rotations and boosts are
\begin{equation}
    \bos r^a_{\ph{s} i} 
    \equiv 
    {}\thalf \ep_{a b c}
    L^{bc}_{\ph{bc} i}
\qand
\bos b^a_{\ph{a} i} 
\equiv {} 
L^{a0}_{\ph{a0} i} \, ,
\label {rotons and boostals}
\end{equation}
and the matrix of gauge fields 
$L_i$ is
\begin{equation}
\begin{split}
L_i = {}&
-i \, \thalf \, \bos r_i \cdot \bos \s \, I
- \thalf \, \bos b_i \cdot \bos \s 
\, \c^5 .
\label {formula for om_i}
\end{split}
\end{equation}
The field strength 
(\ref{field strength of spin connection})
is then
\begin{align}
F_{i k} ={}
\bos [ \p_i +   L_i, 
\p_k +   L_k \bos ] 
={}&
-i \thalf \big[ 
\p_i \bos r_k - \p_k \bos r_i
+   (\bos r_i \times \bos r_k 
- \bos b_i \by \bos b_k ) \big ]
\cdot \bos \s I
\nn\\
{}& - \thalf \big[
\p_i \bos b_k - \p_k \bos b_i 
+  (\bos r_i \by \bos b_k 
+ \bos b_i \by \bos r_k) \big]
\cdot \bos \s \c^5 .
\nn
\end{align}

 \par
  \par
Under a local Lorentz 
transformation 
$D = D(\La(x))$,
these fields transform as
\begin{equation}
\begin{split}
L'^{ab}_{\ph{'ab} i} ={}&
 \La^a_{\ph{a} c}  \La^b_{\ph{b} d}
L^{cd}_{\ph{cd} i}
- {}
\thalf  \tr \lt( D 
\p_i D^{-1}
\bos [ \c^a, \c^b \bos ] \rt)
\\
\p_i + L'_i = {}&
D 
\, \big( 
\p_i  + L_i 
\big) \, D^{-1}
\\
F'_{ik} = {} &
D \, F_{ik} \, D^{-1}
\\
h' = {}& D^{-1\dag} \, h \, D^{-1}
\\
(\b h \b)' ={}&
D \, \b h \b \, D^{\dag} 
\\
\psi'={} & D \, \psi
\label {how they go}
\end{split}
 \end{equation}
and so
the action density $s_L$ 
of the action $S_L$
(\ref{action of the Lorentz connection})
is invariant
\begin{align}
s'_L ={}&
\tr \lt( F'^\dag_{ik} \, h' 
\, F'^{ik} \,  \b h' \b \rt) 
={}
\tr \lt(  
D^{-1 \dag} F^\dag_{ik} 
D^{\dag} \, 
D^{-1 \dag} h D^{-1}
\, D F_{ik} D^{-1} \,
D \b h \b D^{\dag} \rt)
={}
\tr \lt(F^\dag_{ik} \, h 
\, F^{ik} \, \b h \b \rt)
= s_L  
\end{align} 
under local Lorentz transformations
(\ref{Local Lorentz invariance}) as well as
under
general coordinate
transformations
(\ref{General coordinate invariance}).

\par
The squares of the 
time derivatives 
$(\dot L^{ a b }_{\ph{a b} i})^2$ 
and
$(\dot L^{ a 0 }_{\ph{a 0} i})^2$ 
of the gauge fields must
appear with a positive sign
in the action $S_L$ 
(\ref{action of the Lorentz connection})
if the gauge-field action
is to be bounded below.
They will
appear with a positive sign
at low temperatures
if the vector boson $K_i$
in the matrix 
$h = {} i \b \c ^a c_a^{\ph{a} i} K_i $
has 
an average value 
$\la 0 | K_i | 0 \ra = K_{0i}$ in the
low-temperature 
ground state
that is timelike, 
$ K_{0i} \, K_0^{\ph{0} i} 
\simeq - m^2 < 0$.
The matrix $h$, the
vector $K_i$, and
the signs of the 
time derivatives 
$(\dot L^{ a b }_{\ph{a b} i})^2$ 
and
$(\dot L^{ a 0 }_{\ph{a 0} i})^2$ 
are discussed in 
Sec.~\ref{The Matrix h and the Vector K}.
 \par
 The tensor gauge fields 
 $L^{ab}_{\ph{ab} i}$
 have spin 2 (not 3)
 because they 
 are antisymmetric in 
 $a$ and $b$.
\par
An explicit formula
for the matrix $D(\La)$ 
is given in 
Appendix~\ref{The Matrix D app}
along with
derivations of the identities
\begin{equation}
D^\dag \b D ={} \b
\qand
D \b D^\dag ={} \b  .
\label {DbDb}
\end{equation}
\par
These identities
(\ref{DbDb})
imply that the
action $S_L$ with
$h$ replaced by $\b$,
i.e,
the trace
$\tr \big( F^\dag_{ik} \b 
F^{ik} \b \big)$,
is invariant under
local Lorentz transformations. 
But the choice
$ h \to \b $
gives an action 
in which the 
squares of the 
time derivatives
of the tensor gauge fields that gauge
boosts and those that 
gauge rotations 
occur with
opposite signs.
So the trace 
(\ref{action of the Lorentz connection})
which uses the matrix 
$h ={} i \b \c ^a K_a$
may be the only viable choice.
\par
With the abbreviations
\begin{equation}
\begin{split}
\bos R_{ik} ={}&
\p_i \bos r_k - \p_k \bos r_i
+   \, ( \bos r_i \times \bos r_k 
- \bos b_i \by \bos b_k ) 
\\
\bos B_{ik} =&
\p_i \bos b_k - \p_k \bos b_i 
+   \, ( \bos r_i \by \bos b_k 
+ \bos b_i \by \bos r_k ) ,
\end{split}
\end{equation}
the field strength $F_{ik}$ is
$F_{ik} ={} 
- i \thalf \bos R_{ik} 
\cdot \bos \s \ I 
- \thalf \bos B_{ik} 
\cdot \bos \s \, \c^5$, 
and the action $S_L$ is 
\begin{align}
S_L ={}&
- \frac{1}{16 m^2 \l^2} \int \tr
\Big[\big( \bos R_{ik} \cdot \bos \s I 
+i  \bos B_{ik} \cdot \bos \s \c^5
\big) h \, 
\big( \bos R^{ik} \cdot \bos \s I 
- i \bos B^{ik} \cdot \bos \s \c^5
\big) \b h \b
\Big]  \sqrt{g} \, d^4x .
\label {action S_L}
\end{align}

\par
Tensor gauge fields
$L^{ a b }_{\ph{a b} i}$
possess at least 
two other actions
that are invariant
under local Lorentz
transformations
and general
coordinate transformations.
One is succinct
and linear 
\begin{equation}
S_E ={} M^2_E \int 
F^{ab}_{\ph{ab} ik} \,
c_a^{\ph{a} i} \, c_b^{\ph{b} k}
\sqrt{g} 
\, d^4\! x
\label {S_E}
\end{equation}
in the
coefficients
$F^{ab}_{\ph{ab} ik}$
of the field strength
\begin{align}
F_{ik} ={}&
F^{ab}_{\ph{ab} ik}
\, \bos [ 
\gamma_a, \gamma_b \bos ]
=
\p_i L_k - \p_k L_i 
+   \bos [ L_i,L_k \bos]  .
\end{align}
 These coefficients
 \begin{align}
F^{ab}_{\ph{ab} ik}
={}&
\p_i 
L^{ a b }_{\ph{a b} k}
-
\p_k 
L^{ a b }_{\ph{a b} i}
+
L^{b c}_{\ph{b c} \, i}
\,
L^{ a}_{\ph{a} c k}
\label{Fab is}
-
L^{ a c }_{\ph{a c} i}
\,
L^{ b }_{\ph{b} c \, k}   
\nn
\end{align}
resemble 
Riemann's curvature tensor
\begin{equation}
R^k_{\phantom{k} i \ell n} ={}
\p_\ell \Gamma^k_{\ph{k} i n} 
- 
\p_n \Gamma^k_{\ph{k} i \ell } 
+ \Gamma^k_{\ph{k} m \ell} \,
\Gamma^m_{\ph{m} i n }  
- \Gamma^k_{\ph{k} m n} \, 
\Gamma^m_{\ph{m} i \ell}  .
\label {Riemann's curvature tensor}
\end{equation}
But the action $S_E$
(\ref{S_E}) does not
lead to second-order
differential equations
for the gauge fields
$L^{ a b }_{\ph{a b} i}$.

\par
A third invariant action
is based upon 
the scalar 
\begin{align}
F^{ab}_{\ph{ab} ik}
& F_{ab}^{\ph{ab} ik}
 =
{}
\Big( \p_i 
L^{ a b }_{\ph{a b} k}
-
\p_k 
L^{ a b }_{\ph{a b} i}
+
L^{b c}_{\ph{b c} \, i}
\,
L^{ a}_{\ph{a} c k}
-
L^{ a c }_{\ph{a c} i}
\,
L^{ b }_{\ph{b} c \, k}   
\Big)
{}
\Big( \p^i 
L_{ a b }^{\ph{a b} k}
-
\p^k 
L_{ a b }^{\ph{a b} i}
+
L_{b d}^{\ph{b d} \, i}
\,
L_{ a}^{\ph{a} d k}
\label{Fab is}
-
L_{ a d }^{\ph{a d} i}
\,
L_{ b }^{\ph{b} d \, k}   
\Big) .
\nn
\end{align}
It is hermitian and
invariant
under general coordinate
transformations and local
Lorentz transformations, but
the squares of  
its time derivatives 
occur with opposite signs.

\section{The Matrix $h$ and the Vector $K$}
\label{The Matrix h and the Vector K}

\par
The action $S_L$
of the proposed
tensor gauge fields
will be invariant under 
local Lorentz
transformations 
(\ref{how they go})
if the matrix $h$ 
which appears in the trace
(\ref{action of the Lorentz connection})
transforms as
\begin{equation}
h' ={}
D^{-1 \dag} \, h \, D^{-1}
\label {h's transformation law}
\end{equation}
where $D = D(\La(x))$,
$D^\dag \b D ={} \b$
and
$D \b D^\dag ={} \b$ .
\par
The simplest choice is 
the hermitian matrix
\begin{equation}
h = {} 
 i \b \c^a 
 c_a^{\ph{a} i} K_i 
\label {one choice for h}
\end{equation}
in which $K_i$ is 
a real vector
transforming as
\begin{equation}
K'_i(x') ={} 
\frac{\p x^k}{\p x'^i}
\, K_k (x)
\end{equation} 
under general coordinate  
transformations. 
Its action $S_K$
(\ref{S_K})
is simpler than it looks 
since
\begin{equation}
D_i K_k - D_k K_i
={}
\p_i K_k - \p_k K_i .
\label {simpler term}
\end{equation} 
\par
Under a Lorentz
transformation $\La$,
Dirac's gamma matrices
transform as a 4-vector
\begin{equation}
\begin{split}
D(\La) & \c^a D^{-1}(\La)  ={}
\La_b^{\ph{b} a} \, \c^b 
\\
D^{-1}(\La) & \c^a D(\La)  ={}
\La^a_{\ph{a} b} \, \c^b  
\label {Dgamma/D}
\end{split}
\end{equation} 
where $ \La_b^{\ph{b} a}
= \La^{-1 a}_{\ph{-1 a} b}  $.
And since 
the matrix $D(\La)$ leaves 
$\b$ invariant
$ D^{-1 \dag}(\La) \, \b
={} \b \, D(\La)  $
as seen earlier (\ref{DbDb}),
the matrices $\b \c^a$
also transform as a 4-vector
\begin{equation}
\begin{split}
D^{-1 \dag}(\La) \, 
\b & \c^a \, D^{-1}(\La)  ={}
\La_b^{\ph{b} a} \, \b \c^b 
\\
D^{\dag}(\La) \, 
\b & \c^a \, D(\La)  ={}
\La^a_{\ph{a} b} \, \b \c^b .
\label {beta gamma^a is a 4-vector}
\end{split}
\end{equation} 
So $\La$ changes $h'$ to 
\begin{equation}
\begin{split}
h' ={}& i \b \c^a c'^{\ph{a} i}_a
K'_i 
= i \b \c^a 
\La_a^{\ph{a} b} \,
c^{\ph{b} i}_b \, K_i 
={}
i D^{-1 \dag} \b \c^b  D^{-1} \, 
c^{\ph{b} i}_b K_i  
={}
D^{-1 \dag} \, h \, D^{-1}   
\end{split}
\end{equation} 
which satisfies 
(\ref{how they go})
as does $\b h \b$ since
\begin{equation}
\begin{split}
(\b \, h \, \b)' ={}&
\b \, D^{-1 \dag} h D^{-1} \,  \b
= D \, \b h \b \, D^{\dag} .
\end{split} 
\end{equation} 
\par
The squares of the 
time derivatives 
$ \dot L^{ a b }_{\ph{a b} i} $
of the gauge fields must
appear with a positive sign
in the action (\ref{S_L})
if the gauge-field action
is to be bounded below.
They will
appear with a positive sign
at low temperatures
if the vector boson $K_i$
in the matrix 
$h = {} i \b \c ^a c_a^{\ph{a} i} K_i $
has an average value 
$K_{0i} ={}
 \la 0 | K_i | 0 \ra$
in the 
low-temperature vacuum
that is timelike, 
$ K_{0i} \, K^{0i} \simeq - m^2 < 0$.
At low temperatures,
the average value $K_{0i}$
is made 
timelike by the second term
$- \tfourth (K_iK^i + m^2)^2$
in its action $S_K$ (\ref{S_K})
which due to antisymmetry
(\ref{simpler term}) 
may be written
in the simpler form
\begin{equation}
\begin{split}
S_K ={}& 
\int \Big[ - \tfourth
(\p_i K_j - \p_k K_i)
(\p^i K^j - \p^k K^i)
\, 
- \tfourth \xi^2 (K_iK^i + m^2)^2 \Big]
\sqrt{g} \, d^4x  .
\label {S_K redux}
\end{split}
\end{equation}

\par

At low temperatures and
presumably
in the rest frame of the CMB,
the vector $K^i$ has 
the average vacuum value
$ \la 0 | K^i | 0 \ra 
= {} K^i_0
=  m \, \d^i_0 $,
and the average value
of the matrix $h$ 
(\ref{one choice for h})
is
\begin{equation}
\begin{split}
h ={} &
i \b \c ^a 
 c_a^{\ph{a} i} K_{0i}
 =
 i \b \c ^0 
 c_0^{\ph{a} 0} K_{00} 
 ={} - m I .
\end{split}
\end{equation} 
Then in the rest frame 
of the CMB and apart from the fluctuations
$k^i = K^i - K^i_0$, 
the action
$S_L$ (\ref{action S_L})
is
\begin{align}
S_L ={}&
- \frac{1}{16 \l^2} \int \tr
\Big[\big( \bos R_{ik} \cdot \bos \s I 
+i  \bos B_{ik} \cdot \bos \s \c^5
\big) 
\, 
\big( \bos R^{ik} \cdot \bos \s I 
- i \bos B^{ik} \cdot \bos \s \c^5
\big) 
\Big]  \sqrt{g} \, d^4x
\label {the form}\\
={}&
- \frac{1}{4 \l^2} \int \lt(
\bos R_{ik} \cdot \bos R^{ik}
+ \bos B_{ik} \cdot \bos B^{ik} 
\rt) \sqrt{g} \, d^4x  
\nn
\end{align}
in which the squares of the 
time derivatives 
$(\bos {\dot  r_k})^2$
and $(\bos {\dot  b_k})^2$
appear with positive signs
as promised.
So the action $S_L$ is
bounded below in the
rest frame of the CMB.

\par
At low temperatures,
the vector boson $K_i$
fluctuates about its average value,
$K_i (x) = K_{0i} + k_i(x)$.
The fluctuations $k_i(x)$ are
those of a massless vector
field with $k_0(x) = 0$
as discussed in 
Appendix~\ref{The Vector Boson K^i}.
The six gauge fields 
$ L^{ab}_{\ph{ab} i} $
remain massless despite
the nonzero average
value $K^i_0$ of the
vector boson $K^i$.
\par
The ground state 
$|0, \bos v \ra$
of a Lorentz frame
moving at velocity $\bos v$
relative to the CMB is related
to the ground state of the
CMB by a unitary Lorentz
transformation
$|0, \bos v \ra 
= U_{\bos v} | 0 \ra$
that represents matched
(\ref{twinning condition})
Lorentz and general-coordinate transformations. 
The action is invariant
under 
Lorentz and general-coordinate 
transformations
\begin{equation}
U^{-1}_{\bos v} \, S_L \, 
U_{\bos v}
={} S_L .
\end{equation}
So the average value
of the action in the 
state $|0, \bos v \ra$
is the same as in the
state $|0\ra$ in which
the CMB is at rest
\begin{equation}
\begin{split}
\la 0, \bos v |
S_L | 0, \bos v \ra
={}&
\la 0 | U^{-1}_{\bos v}
\, S_L \,
U_{\bos v} | 0 \ra
=
\la 0 | S_L |0 \ra  .
\end{split} 
\end{equation} 
Thus the action is bounded
below in all Lorentz frames.

\par
The matrix $h$ 
takes a simpler form
in two-component
notation as
discussed in
Appendix~\ref{Two-Component Formalism}.

\par
It may be useful here 
to distinguish different 
kinds of symmetry.
One kind is 
an exact symmetry 
of the action and of
the vacuum, 
like that of the group
$SU_c(3)$ of QCD.

\par
A second kind is a
symmetry of the action
but not of the vacuum,
like that of 
SU$_\ell(2) \otimes$U(1)$_Y$
in which the average value
of a component of 
the Higgs field in 
the low-temperature
vacuum
makes the W and Z bosons 
massive.  
Their masses
make their interactions
of short range and therefore
weak. 

\par
A third kind is symmetries 
of the action
that are intrinsically 
reduced by the ground state
of the actual universe.
As described in
Sec.~\ref{Intrinsic Reduction of Symmetry},
the average
values of the tetrads
$c_{ 0 i}^a(x) ={}
\tr ( \rho \, c_{\ph{a} i}^{a}(x) )  $
in the ground state
of the late, 
low-temperature universe
reduce the two independent
symmetries of general coordinate
and local Lorentz invariance
to a single exact symmetry of the vacuum ---
local Lorentz invariance.
Every
local Lorentz transformation 
$ \La^{a}_{\ph{a} b}(x)$
must be 
accompanied by a specific 
general coordinate
transformation
(\ref{special coordinate transformation})
\begin{equation}
\frac{\p x'^i}{\p x^k}
={}
c_a^{\ph{a} i}(x) \, 
\La^{a}_{\ph{a} b}(x) \,
c_{\ph{b} k}^b(x)  
\label{specific coordinate transformation}
\end{equation} 
in order to preserve the
average values of the tetrads.

\par
In the theory sketched in this paper,
the average value 
of the vector boson $K_i$
in the present low-temperature
universe and in the
rest frame of the CMB,
$\la 0 | K_i | 0 \ra ={}
m \, \d_i^0$, 
gives the action $S_L$ 
(\ref{S_L}, \ref{action S_L}) 
of the six gauge fields 
$ L^{ab}_{\ph{ab} i} $
the approximate 
and simpler form
(\ref{the form}).
The six gauge fields 
$ L^{ab}_{\ph{ab} i} $ 
remain massless, and
local Lorentz invariance  
remains exact --- apart
from $\la 0 | K_i | 0 \ra$, the CMB,
and the matter and energy
of the actual universe.
\section{Dirac Action}
\label {Dirac Action}
\par
The explicitly hermitian
Dirac action is
\begin{equation}
\begin{split}
S_D 
={}&
- \thalf \!
\int  \Big[ \ovl \psi \, \c^a
c_a^{\ph{a} i} D_i \psi 
+ \! \big( \ovl \psi \, \c^a
c_a^{\ph{a} i} D_i \psi \big)^\dag \Big]
\sqrt{g} \, d^4x 
\label {hermitian Dirac action}
\end{split} 
\end{equation} 
in which 
the covariant derivative
$D_i \psi$ is
\begin{equation}
D_i \psi ={}
\big(\p_i  + \teighth \,   \,
L^{ a b }_{\ph{a b} i}
\, \bos [ \c_a, \c_b \bos ] 
\big) \psi .
\label {Dirac covariant derivative Sec V}
\end{equation}
To avoid clutter,
I am using a single
Dirac field $\psi$ and
am suppressing the 
gauge bosons $A^\a_{i}$ of
$SU_c(3) \by SU_\ell(2) \by U(1)$.
To include them, one would replace
the single Dirac field $\psi$ by a 
vector $\Psi$ whose components
$\Psi_\a$ would be the 6 leptons
and 18 quark fields.  One would add
the 12 gauge bosons $A^\a_{i}$ of
$SU_c(3) \by SU_\ell(2) \by U(1)$
and their actions. Then the 
covariant derivative of 
the 96-component Dirac field $\Psi$
would be
\begin{equation}
D_i \Psi ={}
\big(\p_i  + \teighth \,   \,
L^{ a b }_{\ph{a b} i}
\, \bos{[} \c_a, \c_b \bos ] 
+ A^\a_{i} t^{\a}
\big) \Psi
\end{equation}
in which the $t^{\a}$'s are the
generators of the Lie algebras of 
$SU_c(3), SU_\ell(2)$ and $U(1)_Y$.

\par
The simplest choice
for $\ovl \psi$ is Dirac's choice
$\ovl \psi = {} \psi^\dag \b$
\begin{equation}
\mathcal{L}_{D}
={} - \psi^\dag  \b  \,  
c_b^{\ph{b} i} \c^b
D_i \psi
\label{alternative Dirac action}
\end{equation}
in which the real $4\by4$ 
hermitian symmetric matrix 
$\b = i \c^0$ obeys
$ D^{\dag}(\La) \, \b \,
D(\La) ={} \b $.
The resulting Dirac action
(\ref{Dirac action})
is then invariant under 
local Lorentz and
general coordinate
transformations. 
\par
Under a
local Lorentz transformation
$ D{(\La)} ={} D(\La(x)) $,
the Dirac field $\psi$,
its covariant derivative $ D_i \psi $
and the tetrads $c_a^{\ph{a} i}$
transform as
\begin{align}
\psi'={} &
D{(\La)} \, \psi
\nn\\
\ovl \psi' ={}&
(\psi^\dag \b)' = {}
\psi^\dag D^{\dag}{(\La)} 
\b 
= \psi^\dag \b \, D^{-1}{(\La)}
\nn\\
(D_i \psi)' ={}&
D{(\La)} \, D_i \psi
\label{the local Lorentz transformation}\\
(c_a^{\ph{a} i})'
={}&
\La_a^{\ph{a} b} \, c_b^{\ph{b}i} 
\nn
\end{align} 
while
$
D{(\La)} \, \c^a \, D^{-1}{(\La)} ={}
\La_b^{\ph{b} a} \, \c^b 
= {} \La^{-1 a}_{\ph{-1 a} b} \c^b $.
Thus the action density 
(\ref{Dirac action})
%
%
is invariant under 
local Lorentz transformations
\begin{align}
\mathcal{L}'_D 
={}&
\lt( \psi^\dag \b \,
\c^a \, c_a^{\ph{a} i}   
D_i \psi \rt)'
={}
\psi^\dag D^{\dag}{(\La)} \, \b  \,  \c^a 
\, D{(\La)} \,
c_c^{\ph{c}i} \La_a^{\ph{a} c}       
\, D_i \psi         
={}
\psi^\dag  \b  \, \c^b \, 
\La^{a}_{\ph{a} b}  
\, \La^{-1 c}_{\ph{-1 c} a} \, c_c^{\ph{c}i}
D_i \psi
={}
\psi^\dag  \b  \, \c^b \,
c_b^{\ph{b} i}          
D_i \psi
= \mathcal{L}_D
\end{align} 
as well as under 
general coordinate transformations.
\par
The explicitly hermitian 
action density is
\begin{align}
\mathcal{L}_{Dh} 
={}&
- \psi^\dag \b 
c_a^{\ph{a} i} 
\c^a \p_i \psi 
- \thalf \psi^\dag \, \b \, \c^a 
\, (\p_i c_a^{\ph{a} i}) \, \psi
\, + \, 
\psi^\dag 
\thalf 
\Big(  c_0^{\ph{0} i} 
\bos r_i \cdot \bos \s \, I 
- c_s^{\ph{s} i} r_i^{\ph{i} s}
\c^5 \Big) 
\psi 
\, - \, \thalf \,  
\ep_{jsk} c_j^{\ph{j}i} b_i^{\ph{i} s} \psi^\dag \s^k \psi  .
\end{align} 
as shown in
Appendix~\ref{Hermitian Form of the Dirac Action app}.
\par
Although the current
that generates the rotational field
$ r^s_i $ is
\begin{equation}
\begin{split}
j_r^{i s} ={}&
\thalf \psi^\dag 
\Big( c_0^{\ph{0} i} \s^s - c_s^{\ph{s} i} 
\c^5 \Big) \psi ,
\end{split}
\end{equation} 
the current that 
generates the boost field
$ b^s_i $ has
no time component
\begin{equation}
j^{is}_b ={}
- \thalf \ep_{jsk} 
c_j^{\ph{j}i}  \psi^\dag 
\s^k \psi .
\end{equation}
So unless the spatial
tetrads are nondiagonal
so that $ c_j^{\ph{j} 0} \ne 0$
for $j = 1, 2$ or 3,
the time components
of the boost bosons
$b_0^{\ph{0} s}$ 
do not occur in the Dirac action,
and 
do not generate  
Coulomb potentials.
\par
These comments apply 
also to 
the gauge fields of groups
larger than the Lorentz group:
unless the spatial
tetrads are nondiagonal,
$ c_a^{\ph{a} 0} \ne 0$
for $a > 0$,
the time components of the 
gauge bosons of the
generators of the 
noncompact directions 
do not appear in the Dirac action
and 
do not generate Coulomb
potentials. 
\bigskip
\par
In the static limit,
the exchange of 
the three massless
tensor gauge fields $\bos r_i$
that gauge rotations
would 
imply that two macroscopic
bodies of $F$ and $F'$ fermions
separated by a distance $r$
would contribute to the energy
a static Coulomb potential
\begin{equation}
K_L(r) = \frac{3 \, F F' f^2}{4 \pi r} .
\label{would have a static potential}
\end{equation}
This potential is positive and repulsive
(between fermions 
and between antifermions).
It violates the weak equivalence
principle
because it depends upon
the number $F$ of fermions
as $F = 3B + L$
and not upon their masses.
\bigskip
\par
The potential $K_L(r)$
changes Newton's potential to
\begin{equation}
K_{NL}(r) = {} - G \, \frac{m m'}{r} 
\lt( 1 + \a  \rt)
\label{K(r)}
\end{equation}
in which the repulsive 
coupling strength $\a$ is 
\begin{equation}
\a = {} - \frac{3 F F'  \l^2}{4 \pi G mm'}
= {} - \frac{3 F F' m_{\text{P}}^2 \l^2}{4 \pi mm'} .
\label{alpha is}
\end{equation}
This force would increase the
need for dark matter and decrease 
the need for dark energy.
Couplings $\a \sim 1$ are
of gravitational strength. 

\par
Experiments~\citep{Harris:2000zz,
Chen:2014oda,
Lee:2020zjt, Tan:2020vpf,
Berge:2017ovy,
Tan:2016vwu,
SQYang2012,Lee:2020zjt,
Adelberger:2009zz,
Geraci:2008hb,Kapner:2006si,
Smullin:2005iv,Hoyle:2004cw,
LongChan2003,Chiaverini:2002cb,
Lee:2020zjt,Hoskins:1985tn,
Williams:2004qba,
Adelberger:2003zx,
Moody:1993ir,
Hoskins:1985tn,
Spero:1980zz,
Schlamminger:2007ht,
Decca:2005qz,Chen:2014oda,
Chiaverini:2002cb,Geraci:2008hb,
LongChan2003,Tu:2007zz,
Yang:2012zzb,Tan:2016vwu,
Fischbach:1999bc} 
put no upper limits on
the masses of tensor gauge fields 
and no lower limits
on their coupling $\l$.

\section{Intrinsic Reduction of Symmetry}
\label {Intrinsic Reduction of Symmetry}
\par
When quantizing a gauge theory,
one picks a gauge.
For general relativity
in flat space,
the usual gauge
choice is to set
the average value 
of the metric $g_{ik}(x)$ 
equal to the
Minkowski metric $\eta$
\begin{equation}
\la 0 | g_{ik}(x) | 0 \ra 
={}
\eta_{ik} 
=
\bpm -1 & 0 & 0 & 0\\
0& 1 & 0 & 0 \\
0& 0 & 1 & 0 \\
0& 0 & 0 & 1 \epm .
\end{equation}
The average vacuum value 
of the metric 
is then quadratic in the
average values 
of Cartan's tetrads 
\begin{equation}
\la 0 | g_{ik}(x) | 0 \ra 
= {}
\la 0 | c_{\ph{a} i}^{a}(x) \, 
\eta_{ab} \, c_{\ph{b} k}^{b}(x) 
| 0 \ra = \eta_{ik} .
\end{equation} 
A further gauge choice 
of Lorentz frame sets
the average vacuum values 
of the tetrads equal to 
Kronecker deltas 
\begin{equation}
\la 0 | c_{\ph{a} i}^{a}(x) | 0 \ra
= {} \d_{\ph{a} i}^{a} .
\label{average vacuum values}
\end{equation} 
\par
The independent
symmetry transformations
of general coordinate 
invariance~(\ref{General coordinate invariance})
\begin{equation}
c'^b_{\phantom{' a} i}(x')  ={}
\frac{\p x^k}{\p x'^i} \, 
c^b_{\phantom{b} k}(x)
\end{equation} 
and local Lorentz 
invariance~(\ref{Local Lorentz 
invariance})
\begin{equation}
c'^a_{\ph{' a} i}(x)  ={}
\La^{a}_{\ph{a} b}(x) \,
c^b_{\phantom{b} i}(x)  
\label{local Lorentz invariance}
\end{equation} 
map a tetrad 
$c^b_{\phantom{b} k}(x)$ 
to
\begin{equation}
\begin{split}
c'^a_{\ph{' a} i}(x')  ={}&  
\La^{a}_{\ph{a} b}(x)  \,
\frac{\p x^k}{\p x'^i} \, 
c^b_{\phantom{b} k}(x) 
\label{joint transformations on tetrads}
\end{split} 
\end{equation} 
\par
So if the average 
vacuum values
(\ref{average vacuum values})
of the tetrads 
$ \la 0 | c_{\ph{a} i}^{a}(x) | 0 \ra
= {} \d_{\ph{a} i}^{a} $
are to be invariant, then
\begin{equation}
\begin{split}
\La^{a}_{\ph{a} b}(x)  \,
\frac{\p x^k}{\p x'^i} \, 
\d^b_{\phantom{b} k} 
=
\La^{a}_{\ph{a} k}(x)  \,
\frac{\p x^k}{\p x'^i} 
= \d^a_{\ph{a} i} .
\label {combined rules}
\end{split}  
\end{equation} 
By multiplying 
this last equation
by $\p x'^i/\p x^\ell$,
we see that
the tetrad values
$ \la 0 | c_{\ph{a} i}^{a}(x) | 0 \ra
= {} \d_{\ph{a} i}^{a} $
will be unchanged 
only if
the general coordinate
transformation $ x \to x'$
and the      
local Lorentz transformation
$\La^{a}_{\ph{a} \ell}(x)$
\begin{equation}
\frac{\p x'^a}{\p x^\ell}
={}
\La^{a}_{\ph{a} \ell}(x)
\quad 
\iff
\quad
\frac{\p x^k}{\p x'^i}
= 
\La^{-1 k}_{\ph{-1 k} i}(x)  
\label {twinning condition}
\end{equation} 
are the same. 
In that case, we have
\begin{equation}
\La^{a}_{\ph{a} k} \,
\frac{\p x^k}{\p x'^i}
={}
\La^{a}_{\ph{a} k} \,
\La^{-1 k}_{\ph{-1 k} i}(x) 
= \d^a_{\ph{k} i} 
\end{equation} 
which maintains
the average values 
of the tetrads in the vacuum 
\begin{equation}
\begin{split}
\la 0 | c'^a_{\ph{' a} i}(x')  | 0 \ra 
={} &
\La^{a}_{\ph{a} b}(x) \,
\frac{\p x^k}{\p x'^i} \, 
\la 0 | c^b_{\phantom{b} k}(x) | 0 \ra
={}
\La^{a}_{\ph{a} b}(x) \,
\frac{\p x^k}{\p x'^i} \, 
\d^b_{\phantom{b} k}
={}
\La^{a}_{\ph{a} k}(x) \,
\La^{-1 k}_{\ph{-1 k} i}(x)
={}
\d^a_{\ph{k} i} .
\end{split} 
\end{equation} 
\par
In a universe
described by a 
density operator $\rho$,
the average values
of the tetrads are traces
\begin{equation}
c_{\ph{a} i}^a(x)
={}
\tr ( \rho \, c_{\ph{a} i}^{a}(x) ) .
\label {average system values}
\end{equation}
By using the tetrad identity
$c_a^{\ph{a} k} 
c^a_{\ph{a} i} = \d^k_i$,
one may show that the 
joint transformations
(\ref{joint transformations on tetrads}) preserve
the average values
(\ref{average system values})
of the tetrads
if the general coordinate transformation
$x \to x'$ is related to the 
local Lorentz transformation 
$\La(x)$ by two tetrads
\begin{equation}
\frac{\p x'^i}{\p x^k}
={}
c_a^{\ph{a} i}(x) \, 
\La^{a}_{\ph{a} b}(x) \,
c_{\ph{b} k}^b(x)  .
 \label{special coordinate transformation}
\end{equation} 
\par
The nonzero average 
values of the tetrads reduce
the two symmetries of the action 
to a single symmetry 
of the ground state of the universe: 
local Lorentz invariance.
This reduction of symmetry
is intrinsic rather than spontaneous
because tetrads intrinsically have
nonzero average values 
$\tr ( \rho \, 
c_{\ph{a} i}^{a}(x) )$. 
\par
The ideas of this section
are independent of the 
existence of the 
tensor gauge fields
proposed in this paper.

\section{Is the gauge group U(2,2)?}
\label {Is the gauge group U(2,2)? sec}
 
 So far we have been assuming that
 the gauge group
of the Dirac field is the Lorentz group
which
acts on the Dirac field as the
direct sum 
 \begin{equation}
 D(\La) ={} D^{(1/2,0)}(\La) 
 \oplus D^{(0,1/2)}( a) .
 \label{direct sum}
 \end{equation}
 Is this a hint of a larger
 symmetry?
These $4\by4$ matrices 
$D(\La) = D(\La(x))$ leave
$\b = i \c^0$ invariant 
\begin{equation}
D^{\dag}(\La) \, \b \,
D^{}(\La) ={} \b
\qand
D(\La) \b D^\dag(\La) ={} \b  .
\label {DbDb redux}
\end{equation}
as noted earlier
(\ref{DbDb})
and transform 
Dirac's gamma matrices 
as a 4-vector 
 \begin{equation}
 D^{\dag}(\La) \, \b \c^a 
 \, D(\La) ={}
\La^a_{\ph{a} b} \, \b \c^b 
 \label{how gammas go under Lamda}
 \end{equation}
 as noted earlier
 (\ref{Dgamma/D} and
 \ref{beta gamma^a is a 4-vector}).
  \par
 The gauge group of
 the Dirac field may be 
 the group of all 
 $4\by4$ complex matrices $U$
 that leave $\b$ invariant
 \begin{equation}
 D^{\dag}(U) \, \b \,
D^{}(U) ={} \b .
\label{U invariance of beta redux}
 \end{equation}
 This group has 16 generators
and is known as 
 U(2,2)~\citep{Witten:20221115} 
 as one may see
 by rotating all four of the 
 matrices of this equation 
 of $\b$ invariance
(\ref{U invariance of beta redux})
 from the $\b = i \c^0$
 direction to the $\c^5$
 direction.  This rotation
 shows that the group
 must leave $\c^5$ invariant
 \begin{equation}
 D^{\dag}(U') \, \c^5 \,
D^{}(U') ={} \c^5
 \label{leaves gamma5 invariant}
 \end{equation} 
 which is the defining 
 equation of U(2,2).
 \par
  The $4\by4$ direct-sum  
 matrices $D(\La)$
(\ref{direct sum}) 
leave $\b$ invariant
(\ref{DbDb redux})
and so form a subgroup
of U(2,2).
\par
 To implement
 U(2,2) gauge symmetry,
 we'll need to extend the
 6 generators 
 $\mathcal{J}^{ a b } = {}
 - \tfourth i 
 [\c^a, \c^b] $
 and 6 gauge fields
 $L^{ a b }_{\ph{a b} i}$
 to 16 generators
 $G^A$
 and 16 gauge fields 
 $L^A_{\ph{A} i}$
 so that
 \begin{equation}
 \begin{split}
 D^{\dag}(U) \b  G^A D(U) 
 ={}&
 \mathscr{D}(U)^{A}_{\ph{A} B} 
 \, \b G^B
 \\
 D^{-1}(U)  G^A D(U) ={}&
 \mathscr{D}(U)^{A}_{\ph{A} B} 
 \, G^B
 \\
 \psi' ={}& D(U) \, \psi
 \\
 D_i \psi ={}&
\big(\p_i + \thalf i \, 
L^A_{\ph{A} i}
\, G_A  
\big) \psi  
\\
(D_i \psi)' ={}&
D (U) D_i \psi
 \label{local U(2,2) transformations}
 \end{split}
 \end{equation}
 in which 
 $\mathscr{D}(U)^{A}_{\ph{A} B}$
 is the $16\by16$ matrix that
 represents $U$ in the adjoint
 representation of $U(2,2)$.
 The Dirac action will
 be invariant under these local
 U(2,2) transformations 
 (\ref{local U(2,2) transformations})
 if Cartan's tetrads are also extended
 from four 4-vectors to 16 4-vectors
 transforming as
 \begin{equation}
 \begin{split}
 e'^{A}_{\ph{'A} i} ={}&
 U^A_{\ph{A} B} 
 \, e^B_{\ph{B} i} 
 \label {U(2,2) rules}
 \end{split}
 \end{equation}
\par
The metric $g_{ik}$
must be invariant  
under U(2,2)
\begin{equation}
g_{ik} ={} e^A_{\ph{A} i} e_{A k}
= e'^A_{\ph{A} i} e'_{A k} 
= g'_{ik} .
\end{equation} 
We have seen (\ref{U(2,2) rules})  
that
$e'^{A}_{\ph{'A} i} ={}
 U^A_{\ph{A} B} 
 \, e^B_{\ph{B} i}$,
 but we can choose how
$e_{A i}$ transforms.
If we choose
\begin{equation}
e'_{A \, k} ={} 
F_A^{\ph{A} C} \, 
e_{C k}
\end{equation}
then to keep
\begin{equation}
\begin{split}
g_{ik} = {}&
e^A_{\ph{A} i} e_{f k}
= e'^f_{\ph{f} i} e'_{f k}
={}
e^h_{\ph{\e} i} \, F^A_{\ph{A} C}
\, \, 
 e_{B k} \,
 U^B_{\ph{B} A}
\end{split} 
\end{equation} 
we need
\begin{equation}
U^A_{\ph{A} B} \, \, 
F_A^{\ph{A} C} = \d^C_B 
\qor
(F^{-1})_B^{\ph{B} A} 
={} U^A_{\ph{A} B} .
\end{equation}
for then
we'd have
\begin{equation}
\begin{split}
g_{ik} = {} &
e^A_{\ph{A} i} e_{A k}
=
e'^A_{\ph{A} i} e'_{A k}
= 
U^A_{\ph{A} B} 
 \, e^B_{\ph{B} i} \,\,
F_A^{\ph{A} C} \, 
e_{C k} 
={}
e^B_{\ph{B} i} \,
\d^C_B  \, e_{C k}
={}
 e^C_{\ph{C} i} 
\,
 e_{C k} = g_{ik} .
\end{split}
\end{equation} 

\par
The real Lie algebras su(2,2) and sl(4,R)
are not isomorphic, but over the 
complex numbers they both
belong to $A_3$ in the Cartan-Weyl classification, so their complexifications 
are isomorphic~\cite{Boyer:20221120}. 
It therefore may make sense to consider
the possibility that
GL(4,R) or GL(4,C) is the gauge group
of the Dirac field.

\section{Cartan's first equation of structure}
\label {Cartan's first equation of structure}
\par
If we add all invariant terms 
of dimension $(\text{mass})^4$ or less
to the action (which is by no means 
required in a non-renormalizable
theory of gravity), then
the covariant derivatives
(\ref{covariant derivative of tetrad})
of the tetrads
\begin{equation}
D_i c^a_{\ph{a} k} ={}
\p_i c^a_{\ph{a} k} +
  \, L_{i \ph{a} b}^{\ph{i} a}
\, c^b_{\ph{a} k} - 
\Gamma^\ell_{\ph{\ell} k i} \, 
c^a_{\ph{a} \ell} 
\label{covariant derivative of tetrad redux}
\end{equation} 
would appear in the action
squared and contracted 
\begin{equation}
\begin{split}
S_C 
={}& -
M^2 \int D_i c^a_{\ph{a} k} 
\, D^i c_a^{\ph{a} k} 
\, \sqrt{g} \, d^4x  
\label{contracted square of covariant derivative redux}
\end{split}
\end{equation}
the coefficient $M^2$
being required because
tetrads are dimensionless.
\par
If the mass $M$
in the action $S_C$ 
is huge, say of the order
of the Planck mass $M_P$,
then the 
equation of motion of
the tensor gauge fields 
would be approximately
the condition
that $S_C$ be stationary
\begin{align}
\d S_C 
={}& -
M^2 \int \lt[
\lt( \d D_i c^a_{\ph{a} k} \rt)
D^i c_a^{\ph{a} k}
+
D_i c^a_{\ph{a} k}
\d D^i c_a^{\ph{a} k} \rt]
\sqrt{g} \, d^4x
\nn
\\
={}& -
2 M^2 \int  
\lt( \d L_{i \ph{a} b}^{\ph{i} a} \rt)
c^b_{\ph{a} k} 
D^i c_a^{\ph{a} k} \sqrt{g} \, d^4x 
= 0
\label {d S_C = 0}
\end{align} 
because the other terms
in the action that contain
the fields 
$L_{i \ph{a} b}^{\ph{i} a}$
 --- namely the action terms 
 $S_D$, 
$S_L$ and $S_K$
(\ref{Dirac action}, \ref{action of the Lorentz connection}, \& \ref{S_K}) ---
lack the huge
coefficient $M^2$.
\par
Thus in the limit 
$M^2  \to \infty$,
the equation
of motion of the tensor gauge fields is
\begin{equation}
0 ={} D_i c^a_{\ph{a} k} 
=
\p_i c^a_{\ph{a} k} +
L_{i \ph{a} b}^{\ph{i} a}
\, c^b_{\ph{a} k} - 
\Gamma^\ell_{\ph{\ell} k i} \, 
c^a_{\ph{a} \ell} 
\label{first equation of structure}
\end{equation} 
which is Cartan's 
first equation of structure
usually derived
from the tetrad postulate
or from the assumption
that the torsion vanishes.
Multiplying it
by $c^{c k}$,
we find that in the  
$M^2  \to \infty$ limit, 
the tensor gauge fields 
approach the spin connection
\begin{equation}
\begin{split}
L_{i }^{\ph{i} ac}
\simeq {}&
\Gamma^\ell_{\ph{\ell} k i} \, 
c^a_{\ph{a} \ell} c^{c k}
+ 
c^a_{\ph{a} k} \, \p_i
c^{c k} 
=
\om_i^{\ph{i} ac}  .
\label {ep L = omega}
\end{split}
\end{equation}

\section{Conclusions}
\label{Conclusions}

The action of
general relativity with fermions 
has two independent symmetries:
invariance under
general coordinate 
transformations 
and invariance under  
local Lorentz transformations.
The action of 
local Lorentz transformations
on Dirac and Lorentz indices
is similar to the action of
noncompact 
internal-symmetry
transformations on
Lie-group indices.
\par
The internal-symmetry 
character of 
local Lorentz invariance 
suggests that it might be
implemented not by
the spin connection
but by
tensor gauge fields 
$L^{ a b }_{\ph{a b} i}$ 
with their own 
Yang-Mills action.
But because the Lorentz group
is noncompact, 
their Yang-Mills action
must be modified by 
a neutral vector field 
$K_i(x)$
whose average value
at low temperatures
is timelike.
This vector boson is massless at low temperatures. 
The vector gauge fields
$L^{ a b }_{\ph{a b} i}$ 
are massless at all temperatures.
\par
The particles of the neutral, gravitationally interacting and massless fields $L^{ a b }_{\ph{a b} i}$  and 
$K_i(x)$ would contribute to the hot dark matter of the Universe. The massive particles radiated by the vector field $K_i$ at high temperatures would contribute at lower temperatures to cold dark matter.
\par
The nonzero
average values
of the tetrads 
reduce the spacetime
symmetries
of the vacuum 
to local 
Lorentz invariance
which can 
be extended to local
U(2,2) invariance.
\par
If the contracted squares of 
the covariant derivatives  
of the tetrads  
multiplied by the square 
of a mass $M$
are added to the action,
then in the limit 
$M^2 \to \infty$,
the equation of motion 
of the tensor gauge fields
is the vanishing of the 
covariant derivatives 
of the tetrads, which is
Cartan's first equation of structure.
In the same limit,
the tensor gauge fields 
approach
the spin connection.

\par

\begin{acknowledgements}
I must thank Charles Boyer, 
Dorothy Burnet,
Bobby Middleton, 
Alain Comtet, 
Peter van Nieuwenhuizen,
Teun van Nuland 
and Edward Witten
for valuable email,
and Michael Grady for helpful
conversations.
\end{acknowledgements}

\appendix

\section{The Matrix D($\Lambda$)}
\label {The Matrix D app}

The matrix 
$D(\La(\bos \th, \bos \l ))$
is
\begin{align}
D(\La(\bos \th, \bos \l ))
={}&
\begin{pmatrix}
D^{(1/2,0)}(\bos{\thet},\bos{ \l }) 
& 0 \\
0 &
D^{(0, 1/2)}(\bos{\thet},\bos{ \l }) 
\end{pmatrix} 
={}
\begin{pmatrix}
e^{\mbox{} - \bos{z} \cdot \bos{\s}}
& 0 \\
0 & 
e^{\mbox{} \bos{z}^* \cdot \bos{\s}} 
\end{pmatrix} 
\label {D(th,lam)}
\end{align} 
in which $\bos z ={} 
\thalf( \bos \l  + i \bos \th 
)$~\citep{Cahillsu2}.
So
\begin{align}
D \b ={}&
\begin{pmatrix}
e^{\mbox{} - \bos{z} \cdot \bos{\s}}
& 0 \\
0 & 
e^{\mbox{} \bos{z}^* \cdot \bos{\s}} 
\end{pmatrix} 
\begin{pmatrix}
0 & 1 \\
1 & 0
\end{pmatrix}
={}
\begin{pmatrix}
0 & e^{-\bos z \cdot \bos \s} \\
e^{\bos z^* \cdot \bos \s} & 0
\end{pmatrix}
={} 
\begin{pmatrix}
0 & 1 \\
1 & 0
\end{pmatrix}
\begin{pmatrix}
e^{\bos z^* \cdot \bos \s} & 0 \\
0 & e^{- \bos z \cdot \bos \s}
\end{pmatrix}
= \b D^{\dag -1}
\end{align}
and
\begin{align}
\b D^{-1} ={}&
\begin{pmatrix}
0 & 1 \\
1 & 0
\end{pmatrix}
\begin{pmatrix}
e^{\bos z \cdot \bos \s} & 0 \\
0 & e^{-\bos z^* \cdot \bos \s}
\end{pmatrix}
={}
\begin{pmatrix}
0 & e^{-\bos z \cdot \bos \s} \\
e^{\bos z \cdot \bos \s} & 0
\end{pmatrix}
={} 
\begin{pmatrix}
e^{- \bos z^* \cdot \bos \s} & 0 \\
0 & e^{\bos z \cdot \bos \s}
\end{pmatrix}
\begin{pmatrix}
0 & 1 \\ 
1 & 0
\end{pmatrix}
= D^\dag \b 
\end{align}
as noted earlier (\ref{DbDb}).

\section{Explicit Form of $S_L$ in Frame of Earth}
\label{Explicit Form of S_L in Frame of Earth app}

The Solar System moves at 
$v = {} 368 \pm 2$ km/s 
relative to the CMB. 
So in the Lorentz frame
of the Earth, 
the average
value of $K^i$ is
$ K^i_{0 v} ={}
\la 0, v | K^i | 0, v \ra 
={} m (1, - \bos v)/
\sqrt{1-v^2} $.
The average value
of the matrix 
$h = {} 
 i \b \c^a 
 c_a^{\ph{a} i} K_i $ 
is then
\begin{equation}
\la 0, E | h | 0, E \ra
={} m \, 
\frac{\bos \s \cdot \bos v \c^5 - I }
{\sqrt{1-v^2}} .
\end{equation} 
Since 
$\b \bos \s \cdot \bos v \c^5 \b = - \bos \s \cdot \bos v \c^5$,
the average value
of $\b h \b$ is
\begin{equation}
\la 0, E | \b h \b | 0, E \ra
={} - m \frac{\bos \s \cdot \bos v \c^5 + I }
{\sqrt{1-v^2}} .
\end{equation}
\par
So at low temperatures
and apart from the 
fluctuations $k_i(x)$,
the action $S_L$
(\ref{action S_L}) is
\begin{align}
S_L ={}&
- \frac{1}{16 m^2 \l^2} \int \tr
\Big[\big( \bos R_{ik} \cdot \bos \s I 
+i  \bos B_{ik} \cdot \bos \s \c^5
\big) h
\,
\big( \bos R^{ik} \cdot \bos \s I 
- i \bos B^{ik} \cdot \bos \s \c^5
\big) \b h \b
\Big]  \sqrt{g} \, d^4x 
\label {<h> S_L}
\nn\\
={}&
\frac{1}{16 \l^2} 
\int \tr
\Big[\big( \bos R_{ik} \cdot \bos \s I 
+i  \bos B_{ik} \cdot \bos \s \c^5
\big) 
\frac{\bos \s \cdot \bos v \c^5 - I }
{1-v^2} 
\,
\big( \bos R^{ik} \cdot \bos \s I 
- i \bos B^{ik} \cdot \bos \s \c^5
\big) (\bos \s \cdot \bos v \c^5 + I )
\Big]  \sqrt{g} \, d^4x .
\end{align}
\par
Since $v \simeq 10^{-3}$,
it's useful to separate
terms according to 
the number of powers
of $v$
\begin{align}
S_L = {}&
- \frac{1}{16 \l^2} \int \tr
\Big[\big( \bos R_{ik} \cdot \bos \s I 
+i  \bos B_{ik} \cdot \bos \s \c^5
\big) 
\,
\big( \bos R^{ik} \cdot \bos \s I 
- i \bos B^{ik} \cdot \bos \s \c^5
\big) 
\Big]  
\frac{\sqrt{g} \, d^4x}
{1-v^2} 
\label {first}
\\
{}&-
\frac{1}{16 \l^2} \int \tr
\Big[\big( \bos R_{ik} \cdot \bos \s I 
+i  \bos B_{ik} \cdot \bos \s \c^5
\big) 
\, 
\big( \bos R^{ik} \cdot \bos \s I 
- i \bos B^{ik} \cdot \bos \s \c^5
\big) \bos \s \cdot \bos v \c^5 
\Big]  \sqrt{g} \, d^4x 
\label{second}\\
{}&+
\frac{1}{16 \l^2} \int \tr
\Big[\big( \bos R_{ik} \cdot \bos \s I 
+i  \bos B_{ik} \cdot \bos \s \c^5
\big) 
\bos \s \cdot \bos v \c^5 
\,
\big( \bos R^{ik} \cdot \bos \s I 
- i \bos B^{ik} \cdot \bos \s \c^5
\big) 
\Big]  \sqrt{g} \, d^4x   
\label{third}\\
{}&+
\frac{1}{16 \l^2} \int \tr
\Big[\big( \bos R_{ik} \cdot \bos \s I 
+i  \bos B_{ik} \cdot \bos \s \c^5
\big) 
\bos \s \cdot \bos v \c^5 
\,
\big( \bos R^{ik} \cdot \bos \s I 
- i \bos B^{ik} \cdot \bos \s \c^5
\big) 
\bos \s \cdot \bos v \c^5
\Big]  
\label{fourth}
\end{align}
in which the first integral
contains terms of order zero and two
in $v$.
\par
Terms with an odd number
of $\c^5$'s cancel.
The first term
(\ref{first}) is thus
\begin{equation}
I_1 ={}
- \frac{1}{4 \l^2} \int \lt(
\bos R_{ik} \cdot \bos R^{ik}
+ \bos B_{ik} \cdot \bos B^{ik} 
\rt) 
\frac{\sqrt{g} \, d^4x}
{1-v^2} .
\end{equation} 
The second
(\ref{second})
and third (\ref{third})
integrals involve a
commutator
\begin{align}
I_2 + I_3 ={}&
\frac{1}{16 \l^2} \int \tr
\Big\{\Big[ 
\big( \bos R_{ik} \cdot \bos \s I 
- i  \bos B_{ik} \cdot \bos \s \c^5
\big) \, ,
\,
\big( \bos R^{ik} \cdot \bos \s I 
+ i \bos B^{ik} \cdot \bos \s \c^5
\big) \Big]
\bos \s \cdot \bos v \c^5
\Big\}  \sqrt{g} \, d^4x  
\nn
\\
={}&
\frac{i}{8 \l^2} \int \tr
\Big\{ 
\Big[ \bos R_{ik} \cdot \bos \s I ,
\bos B^{ik} \cdot \bos \s \Big] \,
\bos \s \cdot \bos v 
\Big\}  \sqrt{g} \, d^4x 
={} -
\frac{1}{\l^2} \int 
\bos R_{ik} \by  
\bos B^{ik} \cdot \bos v  
 \sqrt{g} \, d^4x .
\end{align} 
The fourth term 
(\ref{fourth})
is quadratic in $v$
\begin{align}
I_4 ={}&
\frac{1}{16 \l^2} \int \tr
\Big[\big( \bos R_{ik} \cdot \bos \s I 
+i  \bos B_{ik} \cdot \bos \s \c^5
\big) 
\bos \s \cdot \bos v \c^5 
\,
\big( \bos R^{ik} \cdot \bos \s I 
- i \bos B^{ik} \cdot \bos \s \c^5
\big) 
\bos \s \cdot \bos v \c^5
\Big]  \sqrt{g} \, d^4x 
\nn\\
={}&
\frac{1}{16 \l^2} \int \tr
\Big[\bos R_{ik} \cdot \bos \s \,
\bos \s \cdot \bos v  \,
\bos R^{ik} \cdot \bos \s  
\, \bos \s \cdot \bos v 
\, + \, 
\bos B_{ik} \cdot \bos \s 
\, \bos \s \cdot \bos v \,
\bos B^{ik} \cdot \bos \s 
\, \bos \s \cdot \bos v
\Big]  \sqrt{g} \, d^4x .
\\
={}&
\frac{1}{4 \l^2} \int 
\Big[
2 \bos R_{ik} \cdot \bos v
\, \bos R^{ik} \cdot \bos v
-
\bos R_{ik} \cdot 
\bos R^{ik} \,
\bos v \cdot \bos v
\, + \, 
2 \bos B_{ik} \cdot \bos v
\, \bos B^{ik} \cdot \bos v
-
\bos B_{ik} \cdot 
\bos B^{ik} \,
\bos v \cdot \bos v
\Big]  \sqrt{g} \, d^4x .
\nn
\end{align} 

\par
The action $S_L$ in
the frame of the Earth
is the sum
\begin{equation}
S_L = {} I_1 + I_2 + I_3 + I_4 .
\end{equation}

\section{The Vector Boson $K^i$}
\label {The Vector Boson K^i}

At high temperatures,
the Lagrange density 
\begin{align}
L_K ={}& - \tfourth
(\p_i K_j - \p_j K_i)
(\p^i K^j - \p^j K^i)
\,
- \, \tfourth \xi^2 m^4 - \thalf \xi^2 m^2 K_i K^i
- \tfourth \xi^2 K_i K^i K_j K^j 
\nn
\end{align} 
of the action  $S_K$
(\ref{S_K redux}) describes
a spin-one 
vector boson $K^i$
that of mass $\xi m$.
These massive particles 
of the boson $K_i$ 
would have contributed to
cold dark matter 
at temperatures 
$ T \ll \xi m$.
\par
At the low temperatures 
of the present universe,
the same Lagrange density
describes a vector boson
$K^i$ whose average value
$\la 0 | K^i(x) | 0 \ra$ 
is timelike;
in some Lorentz frame,
presumably 
the rest frame of the CMB,
the vector 
$K^i(x) ={} m \d^i_0 + k^i(x)$ 
has average values
$ \la 0 | K^0(x) | 0 \ra ={} m$
and 
$ \la 0 | k^i(x)  | 0 \ra= 0  $
which make 
$\la 0 | K^i(x) | 0 \ra$ timelike.
Its small fluctuations $k^i(x)$
are those of a massless
vector $\bos k(x)$ 
described by the
Lagrange density 
\begin{equation}
\begin{split}
L_k ={}& 
- \tfourth
(\p_i k_j - \p_j k_i)
(\p^i k^j - \p^j k^i)
\, - \, \tfourth \xi^2 m^4
- \thalf \xi^2 m^2 
\big(\bos k^2 - (m+k^0)^2\big) 
\,
- \, \tfourth \xi^2
\big(
\bos k^2 - (m+k^0)^2
\big)^2
\end{split} 
\end{equation} 
or
\begin{align}
L_k ={}& 
- \tfourth
(\p_i k_j - \p_j k_i)
(\p^i k^j - \p^j k^i)
\, - \, \tfourth \xi^2 m^4
- \thalf \xi^2m^2 
\big(\bos k^2 - m^2
- 2m k^0 - k^{02}
\big) 
\\
{}&
-\tfourth \xi^2 (\bos k^2)^2
+ \thalf \xi^2 \bos k^2
(m^2 + 2mk^0 + k^{02})
\, 
- \, \tfourth \xi^2 \big(
m^4 + 4 m^3 k^0
+ 6 m^2 k^{02}
+ 4 m k^{03}
+ k^{04} \big)  .
\nn 
\end{align} 
These massless particles 
would contribute 
hot dark matter at 
$ T \ll \xi m$.
\par
Combining terms, we get 
\begin{align}
L_k ={}& 
- \tfourth
(\p_i k_j - \p_j k_i)
(\p^i k^j - \p^j k^i)
\, - \, \tfourth \xi^2 m^4 + \thalf \xi^2 m^4
- \tfourth \xi^2 m^4
+ \xi^2 m^3 k^0 - m^3 \xi^2 k^0
\\
{}&
- \thalf \xi^2 m^2 \bos k^2 
+ \thalf \xi^2 m^2 \bos k^2 
+ \thalf \xi^2 m^2 k^{02}
- \threehalves \xi^2 m^2 k^{02}
\, + \, m \xi^2 \bos k^2 k^0
- m \xi^2 k^{03}
- \tfourth \xi^2 (\bos k^2)^2 
+ \thalf \xi^2 \bos k^2 k^{02} 
- \tfourth \xi^2 k^{04}  
\nn
\end{align} 
or
\begin{align}
L_k ={}&
- \tfourth
(\p_i k_j - \p_j k_i)
(\p^i k^j - \p^j k^i)
\,
- \, \xi^2 m^2 k^{02} + \xi^2 m \bos k^2 k^0 
- \xi^2 m k^{03}
\, - \, \tfourth \xi^2 (\bos k^2)^2
+ \thalf \xi^2 \bos k^2 k^{02} 
- \tfourth \xi^2 k^{04} .
\end{align} 
\par
The quadratic part of 
$L_k$ is
\begin{equation}
\begin{split}
L_{k2} ={}& 
- \tfourth
(\p_i k_j - \p_j k_i)
(\p^i k^j - \p^j k^i)
- \xi^2 m^2 k^{02}  .
\end{split}
\end{equation} 
\par
The linear Euler-Lagrange
equations for $k^i(x)$ are
\begin{equation}
\p_j \big( \p^j k^i - \p^i k^j \big)
={} - 2 \xi^2 m^2 \, k^0 \, \d^i_0  
\end{equation} 
or
\begin{equation}
\begin{split}
\triangle k^0 
+ \grad \cdot \dot{\bos k} 
={} - 2 \xi^2 m^2 k^0   
\label {first pair of equations}
\end{split}
\end{equation}
and
\begin{equation}
- \ddot {\bos k} + \triangle \bos k
- \grad \big( \dot k^0 
+ \grad \cdot \bos k \big)
= {} 0  .
\end{equation} 
One solution is 
$- \triangle k^0= 2 m^2 k^0$,
$\grad \cdot \bos k = 0$, 
$ \dot k^0 = 0$, and
$ \p_j\p^j \bos k ={} 0$. 
\begin{align}
\end{align}

\section{Two-Component Formalism}
\label{Two-Component Formalism}
Dirac's formalism is economical,
but the two-component formalism
is better suited to a discussion
of a new interpretation of
the matrix $h$. 
\par
Since the Dirac-Lorentz matrix
(\ref{D(th,lam)}) is 
block diagonal
\begin{align}
D(\bos \th,\bos \l) {}&
\begin{pmatrix}
D^{(1/2,0)}(\bos{\thet},\bos{\lambda }) 
& 0 \\
0 &
D^{(0, 1/2)}(\bos{\thet},\bos{\lambda }) 
\end{pmatrix} 
={}
\begin{pmatrix}
e^{\mbox{} - \bos{z} \cdot \bos{\s}}
& 0 \\
0 & 
e^{\mbox{} \bos{z}^* \cdot \bos{\s}} 
\end{pmatrix} 
\equiv
\begin{pmatrix}
D_\ell
& 0 \\
0 & 
D_r
\end{pmatrix} ,
\label {D(th,lam) in text}
\end{align} 
the matrix $h$ 
\begin{equation}
h ={}
\begin{pmatrix}
h_\ell
& 0 \\
0 & 
h_r
\end{pmatrix}
\end{equation} 
and its
transformation law
$ h' = D^{-1 \dag} \, h \, D^{-1} $
(\ref{h's transformation law})
are block diagonal
\begin{equation}
\begin{split}
\begin{pmatrix}
h'_\ell
& 0 \\
0 & 
h'_r
\end{pmatrix}
={}&
\begin{pmatrix}
D^{-1\dag}_\ell h_\ell D^{-1}_\ell
& 0 \\
0 & 
D^{-1\dag}_r h_r D^{-1}_r 
\end{pmatrix}  .
\end{split} 
\end{equation}
The matrix $\b h \b$
is just $h$ with left and
right interchanged
\begin{equation}
\b h \b ={} 
\begin{pmatrix}
h_r
& 0 \\
0 & 
h_\ell
\end{pmatrix} .
\end{equation} 
\par
The transformation law
$\psi' ={} D \, \psi$
of a Dirac
field 
\begin{equation}
\psi ={} 
\begin{pmatrix}
\ell \\ r
\end{pmatrix} 
\end{equation}
is  
\begin{equation}
\psi' ={} 
\begin{pmatrix}
\ell' \\ r'
\end{pmatrix} 
=
\begin{pmatrix}
D_\ell \, \ell \\ 
D_r \, r
\end{pmatrix} .
\end{equation}
So the combinations 
$ \ell^\dag h_\ell \ell $
and
$ r^\dag h_r r $
are invariant
\begin{equation}
\begin{split}
(\ell^\dag h_\ell \ell)'
={}&
\ell^\dag D_\ell^\dag 
D^{-1\dag}_\ell h_\ell D^{-1}_\ell
D_\ell \ell
=
\ell^\dag h_\ell \ell
\\
(r^\dag h_r r)'
={}&
r^\dag D_r^\dag
D^{-1\dag}_r h_r D^{-1}_r
D_rr
=
r^\dag h_r r
\end{split} 
\end{equation} 
much as contracted 
tensors are invariant
\begin{equation}
(X^i g_{ik} Y^k)'
={}
X'^i g'_{ik} Y'^k
=
X^i g_{ik} Y^k .
\end{equation} 
We now see that 
the $2\by2$ matrices
$h_\ell$ and $h_r$
do for spinor indices what
$g_{ik}$ does for tensor
indices.
\par
Now if 
$D_\ell = e^{- \bos z \cdot \bos \s/2}$,
then 
$D_r = e^{\bos z^* \cdot \bos \s/2}$,
and so
$
D_r = D^{\dag -1}_\ell$.  
So we can set
$
h_r ={} h_\ell^{-1 \dag}$. 
\par
Thus in two-component notation,
the gauge-field action $S_L$
(\ref{action S_L}) is
\begin{equation}
\begin{split}
S_L ={}&
- \frac{1}{16 m^2} \int 
\Big\{ \tr
\Big[\big( \bos R_{ik}  
+ i  \bos B_{ik} \big) 
\cdot \bos \s 
\,  h_\ell
\,
\big( \bos R^{ik} 
- i \bos B^{ik} \big)
\cdot \bos \s 
\, h_\ell^{-1 \dag}
\Big]  
\\
{}&
\qquad \qquad +
\tr
\Big[\big( \bos R_{ik}  
+i  \bos B_{ik} \big)
\cdot \bos \s 
\, h_\ell^{-1 \dag}
\,
\big( \bos R^{ik} 
+ i \bos B^{ik} \big) 
\cdot \bos \s 
\, h_\ell
\Big]  
\Big\}
\sqrt{g} \, d^4x .
\label {action S_L redux}
\end{split}
\end{equation}
\par
Since $h'_\ell = {}
D^{-1 \dag}_\ell h_\ell D^{-1}_\ell $,
we may choose $h_\ell$
to be hermitian
$h^\dag_\ell ={} h_\ell$,
which implies that 
the diagonal form of
$h_\ell$ is just two real 
numbers.
The most general $2\by2$
hermitian matrix is a linear
combination of the Pauli
matices $\bos \s$ and the
identity matrix $I$.
Under a Lorentz transformation
$\La$, the 4-vector 
$s^a_\ell \equiv (-I,\bos{\s})$
transforms as
\begin{equation}
D_\ell^\dagger (\La) \, 
s^a_\ell \, D_\ell(\La)
= \La^a_{\;\; b} \, s^b_\ell 
\label {(-I,sig) under finite Lor trans}
\end{equation}
while the 4-vector 
$s^a_r \equiv (I,\bos{\s})$
transforms as
\begin{equation}
D_r^\dagger (\La) \, 
s^a_r \, D_r(\La)
= \La^a_{\;\; b} \, s^b_r .
\label {(I,sig) under finite Lor trans}
\end{equation}
The $4\by4$ matrix $h$
(\ref{one choice for h}) is
\begin{equation}
h = {} 
 i \b \c ^a 
 c_a^{\ph{a} i} K_i 
 =  i \b \c ^a K_a 
 = \begin{pmatrix}
 s^a_\ell K_a & 0 \\
 0 & s^a_r K_a
 \end{pmatrix}.
\label {two-component choice for h}
\end{equation}
Under a local Lorentz transformation
$ \La = \La(x)$,
the vector field $K_a(x)$ goes to
\begin{equation}
K'_a(x) ={} 
U(\La) K_a(x) U^{-1}(\La)
=
\La^{\ph{a} b}_a
K_b(x) .
\label{SW 5.1.6}
\end{equation} 
So the explicitly
$(\thalf,0) \oplus (0, \thalf)$
version of $h$ goes as
\begin{equation}
\begin{split}
h' ={}&
 \begin{pmatrix}
 s^a_\ell \La_a^{\ph{a} b} K_b & 0 \\
 0 & s^a_r \La_a^{\ph{a} b} K_b
 \end{pmatrix}
  =
D^{-1 \dag} h D^{-1}
\,
 ={} \,
\begin{pmatrix}
D^{-1 \dag}_\ell s^b_\ell D^{-1}_\ell K_b 
& 0 
\\
 0 & D^{-1 \dag}_r s^b_r D^{-1}_r
 K_b
\end{pmatrix}  
\end{split}  
\end{equation} 
since
\begin{equation}
\begin{split}
D^{-1 \dag}_\ell s^b_\ell D^{-1}_\ell K_b
={}&
s^a_\ell \La^{-1 b}_{\ph{-1 b} a} K_b
= s^a_\ell \La_a^{\ph{a} b} K_b
\\
D^{-1 \dag}_r s^b_r D^{-1}_r K_b
={}&
s^a_r \La^{-1 b}_{\ph{-1 b} a} K_b
= s^a_r \La_a^{\ph{a} b} K_b .
\end{split} 
\end{equation} 

\section{Hermitian Form of the Dirac Action}
\label {Hermitian Form of the Dirac Action app}

\par
The fully expanded
covariant derivative
$D_i\psi$ is
\begin{equation}
\begin{split}
D_i\psi ={}&
\lt( \p_i 
-i \, \thalf \,   \,\bos r_i \cdot \bos \s \, I
- \thalf \,   \, \bos b_i \cdot \bos \s 
\, \c^5 \rt) \psi .
\end{split} 
\end{equation} 
If $\ovl \psi ={} \psi^\dag \b$,
then $ - \ovl \psi \, \c^a 
\, c_a^{\ph{a} i} 
D_i \psi $ is
\begin{align}
- \psi^\dag \b \, \c^a 
\, c_a^{\ph{a} i} D_i \psi 
&={}
 - \psi^\dag \b \, \c^a \,
c_a^{\ph{a} i}  
(\p_i \psi ) 
\, 
- \, \psi^\dag \b \, \c^a \,
c_a^{\ph{a} i} 
( - i \thalf \,  
\, \bos r_i \cdot \bos \s \, I)
\psi 
\,
- \, \psi^\dag \b \, \c^a \, c_a^{\ph{a} i} 
(- \thalf \,   \, \bos b_i \cdot \bos \s 
\, \c^5) 
\psi 
\nn\\
&={}
 -  \psi^\dag \b \, 
\c^a  \, c_a^{\ph{a} i} 
(\p_i \psi ) 
\, 
+ \, \thalf \, i \, \psi^\dag \b
\, \c^a \, c_a^{\ph{a} i} 
\,   \, \bos r_i \cdot \bos \s \, I
\psi 
\,
+ \, \thalf \,  
\psi^\dag \b \, \c^a \, c_a^{\ph{a} i} 
\,   \, \bos b_i \cdot \bos \s 
\, \c^5
\psi  .
\label{beta case}
\end{align}  
Its adjoint 
$ [ - \ovl \psi c_a^{\ph{a} i} 
\c^a D_i \psi  ]^\dag $
is
\begin{align}
[ - \psi^\dag \b \, \c^a \,
c_a^{\ph{a} i}  D_i \psi & ]^\dag
={} 
\big[  - \psi^\dag \b 
\, \c^a \, c_a^{\ph{a} i}  
(\p_i \psi )  \big]^\dag
\, + \, 
\big[ 
\thalf i \psi^\dag \b \, \c^a \,
c_a^{\ph{a} i} 
\,   \, \bos r_i \cdot \bos \s \, I
\psi  \big]^\dag
\,
+ \,  
\big[
\thalf \, \psi^\dag \b \, \c^a
\, c_a^{\ph{a} i} \,   \,
\bos b_i \cdot \bos \s 
\, \c^5
\psi  \big]^\dag
\nn\\
{}& \equiv 
A_1 + A_2 + A_3 .
\end{align}  
Now $\b \, \c^a = {} i \c^0 \c^a$
is antihermitian,
so the first term is
\begin{equation}
\begin{split}
A_1 ={}& \big[  - \psi^\dag 
\b \, \c^a
c_a^{\ph{a} i}  
(\p_i \psi )  \big]^\dag
= {}
( \p_i \psi^\dag )
\b \, \c^a \, c_a^{\ph{a} i}  
 \, \psi 
={}
- \psi^\dag \, \b \, \c^a 
\, (\p_i c_a^{\ph{a} i}) \, \psi
- \psi^\dag \, \b \,  \c^a \, 
c_a^{\ph{a} i} \, \p_i \psi .
\end{split} 
\end{equation} 
The second term is
\begin{equation}
\begin{split}
A_2 ={}&
\big[ 
\thalf i \psi^\dag \, \b \, \c^a
\, c_a^{\ph{a} i} 
\,   \, \bos r_i \cdot \bos \s \, I    
\psi  \big]^\dag
={} \,
\thalf \, i \, 
\psi^\dag
\,   \, \bos r_i \cdot \bos \s \, I 
\, \b \, \c^a  
\, c_a^{\ph{a} i} \, \psi  
={}
\thalf \, i \, 
\psi^\dag \, \b \,   \,
\bos r_i \cdot \bos \s \, I 
\, \c^a  
\, c_a^{\ph{a} i} \, \psi  .
\end{split}
\end{equation}
The third term is
\begin{align}
A_3 ={}&
\big[
\thalf \, \psi^\dag \b \, \c^a
\, c_a^{\ph{a} i} \,   \, 
\bos b_i \cdot \bos \s 
\, \c^5
\psi  \big]^\dag
={} 
- \thalf \psi^\dag \, \c^5 
\,   \,
\bos b_i \cdot \bos \s  
\, \b \, \c^a \, c_a^{\ph{a} i} \,
\psi 
={}
-\thalf \psi^\dag \b  \,
\bos b_i \cdot \bos \s 
\, \c^a \, c_a^{\ph{a} i} \,
\c^5 \, \psi .
\end{align} 
\par
Since 
$
[ \s^s, \c^j ] ={} 
2 i \, \ep_{sjk}\c^k $,
we have
\begin{equation}
\s^s \, \c^j ={} 
\c^j \, \s^s + 2 i \, 
\ep_{sjk} \c^k .
\end{equation}
So the $\bos r $ terms are
\begin{align}
\tfourth \, i \, \psi^\dag \b 
\, \c^a \, c_a^{\ph{a} i} 
\,   \, \bos r_i \cdot \bos \s \, I
\psi 
+ &
\tfourth \, i \, 
\psi^\dag \, \b \,   \,
\bos r_i \cdot \bos \s \, I 
\, \c^a  
\, c_a^{\ph{a} i} \, \psi  
={}
\tfourth \,  \psi^\dag 
\Big( 2 c_0^{\ph{0} i} 
\,   \, \bos r_i \cdot \bos \s \, I 
+ \b \, c_j^{\ph{j} i} 
\,   \, r_i^{\ph{i} s}
\{ i\c^j, \s^s \} \Big)
\psi
\nn\\
={}&
\thalf \psi^\dag 
\Big( c_0^{\ph{0} i} 
\,   \, \bos r_i \cdot \bos \s \, I 
+ \b \, c_s^{\ph{s} i} 
\,   \, r_i^{\ph{i} s}
\c^5 \b \Big)
\psi
={} 
\thalf
\psi^\dag 
\Big( c_0^{\ph{0} i} 
\,   \, \bos r_i \cdot \bos \s \, I 
- c_s^{\ph{s} i} \,   \, r_i^{\ph{i} s}
\c^5 \Big) 
\psi  ,
\end{align} 
while the $\bos b$ terms are
\begin{align}
\thalf \,  
\psi^\dag \b & \, \c^a \, c_a^{\ph{a} i} 
\,   \, \bos b_i \cdot \bos \s 
\, \c^5
\psi
-
\thalf \psi^\dag \b \,   \, 
\bos b_i \cdot \bos \s 
\, \c^a \, c_a^{\ph{a} i} \,
\c^5 \, \psi
={}
i \, \ep_{tsk} \, 
c_t^{\ph{t} i} \,   \,
b_i^{\ph{i} s} \,
\psi^\dag \, \b \, 
 \c^k \, \c^5 \, \psi .
\end{align}
\par
The hermitian Dirac action 
density then is
\begin{align}
\mathcal{L}_{Dh} 
={}&
- \psi^\dag \b 
c_a^{\ph{a} i} 
\c^a \p_i \psi 
- \thalf \psi^\dag \, \b \, \c^a 
\, (\p_i c_a^{\ph{a} i}) \, \psi
{} \, + \,
\psi^\dag 
\thalf \,   \, \Big(  c_0^{\ph{0} i} 
\bos r_i \cdot \bos \s \, I 
- c_s^{\ph{s} i} r_i^{\ph{i} s}
\c^5 \Big) 
\psi 
\,
- \, \thalf \ep_{jsk} \,   \, c_j^{\ph{j}i} b_i^{\ph{i} s} \psi^\dag \s^k \psi .
\end{align}

\bibliography{physics}
 
\end{document}